\documentclass[letterpaper,twocolumn,10pt]{article}
\usepackage{usenix,epsfig}

\begin{document}

\date{}

\title{Consideration for effectively handling parallel 
workloads on public cloud system}

\author{
{\rm Kazuichi Oe}\\
National Institute of Infomatics \\
Japan \\
koe@nii.ac.jp
} 

\maketitle


\subsection*{Abstract}

We retrieved and analyzed parallel storage workloads of 
the FUJITSU K5 cloud service to clarify how to build 
cost-effective hybrid storage systems. 
A hybrid storage system consists of fast but low-capacity 
tier (first tier) and slow but high-capacity tier (second tier). 
And, it typically consists of either SSDs and HDDs or NVMs and 
SSDs. 
As a result, we found that 1) regions for first tier should 
be assigned only if a workload includes large number of IO 
accesses for a whole day, 2) the regions that include a large 
number of IO accesses should be dynamically chosen and moved 
from second tier to first tier for a short interval, and 3) if 
a cache hit ratio is regularly low, use of the cache for the 
workload should be cancelled, and the whole workload region 
should be assigned to the region for first tier.
These workloads already have been released from the SNIA web 
site.

\section{Introduction}

\label{intro}

Cloud services are rapidly spreading because they can speed up 
software development for users, reduce a user's operational costs, 
and enable agile development. 
These services are built on virtual machine (VM) environments, 
and these VMs often compete for input/output (IO) performance 
\cite{IC2E_2016}. 
It is important for us to improve the cost-performance of 
cloud services by resolving the conflict of IO performance. 
The FUJITSU K5 cloud service \cite{FUJITSU_K5}
\footnote{The current name is ``FUJITSU Cloud Service for OSS''}
is supported by 
more than 13,000 servers and 640 business systems and is built 
using the OpenStack \cite{OpenStack} and VMware platforms. 
Customer's applications might be executed in parallel on the 
same VMware platform.
And, we installed a capture system in the TCP/IP network between 
servers running on hypervisor and storage systems in a K5 data 
center in Japan \cite{WANC-2019} and retrieved the parallel workloads, 
which consisted of 3088 virtual storages, by using the capture system.
These workloads have already been released from the IO trace web-site 
of the Storage Networking Industry Association (SNIA) 
\cite{SNIA-IO-TRACE}. 
Our goal was to clarify how to build cost-effective hybrid storage 
systems for cloud service platform. 
A hybrid storage system consists of fast but low-capacity 
tier (first tier) and slow but high-capacity tier (second tier). 
Of cource, the cost for first tier is much higher than that for 
second tier. 
And, it typically consists of either solid state drives (SSDs) 
and hard disk drives (HDDs) or non-volatile memories (NVMs) and 
SSDs.  
Therefore, we analyzed these parallel workloads from the viewpoint 
of both temporal and spatial access locality. 
We first examined the bias of the number of IO accesses among the 
3088 workloads and found that only 38 workloads included 50 \% of 
all IO accesses. 
We then analyzed these 38 workloads because of easy and quick 
analysis. 
For temporal access locality, we classified these workloads 
into two IO access patterns when we analyzed them over the course 
of a day. 
One was that the number of IO accesses is high only for a specific 
time, and the other is that the number of IO accesses is almost 
stable. 
When we analyzed them on a daily basis, the number of IO accesses 
for each workload often varied every day. 
For both temporal and spatial access locality, we executed the 
sim-ideal \cite{SIM-IDEAL} cache simulator to investigate page-level 
access regularities. As a result, these cache hit ratios varied with 
the workload, but almost half of the workloads had low cache hit 
ratios because these workloads included few page-level regularities. 
Moreover, we found that almost all workloads concentrated IO accesses 
on a narrow region. 
This narrow region was drastically bigger than the page unit and moved 
in a short interval. Automated Tiered Storage with Fast Memory and 
Slow Flash Storage (ATSMF) \cite{IEICE-D-12} may effectively handle 
these concentrated IO accesses because its replacement algorithm 
may be fit for the concentration. 
Therefore, we had the following ideas for building a 
cost-effective hybrid-storage system. 
A preferable hybrid storage system should dynamically optimized across 
workloads by using the following ideas because multiple workloads with 
different characteristics were running in parallel. 

\begin{itemize}
\item Assign regions for first tier only if a workload includes a 
large number of IO accesses for a whole day. 
\item Dynamically choose the regions that include a large number of 
IO accesses and move them from second tier to first tier for a short 
interval. 
\item Check the cache hit ratio of each workload. If a cache hit ratio 
is regularly low, the use of the cache for the workload should be 
cancelled, and the whole workload region should be assigned to the 
region for first tier. 
If a cache hit ratio increases according to the increase of the 
cache size, the convergence point for the cache hit ratio should be 
searched for, and the corresponding cache size should be set. 
If the cache hit ratios differ among the candidate cache replacement 
algorithms, a more perferable algorithm should be chosen. 
\item Use the feature for concentrated IO accesses on a narrow 
region. ATSMF is one candidate system. 
\end{itemize}

\section{How to retrieve trace logs from FUJITSU 
K5 cloud service}

\label{fj_k5}

\begin{figure}[tb]
\begin{center}
\includegraphics[width=6.5cm]{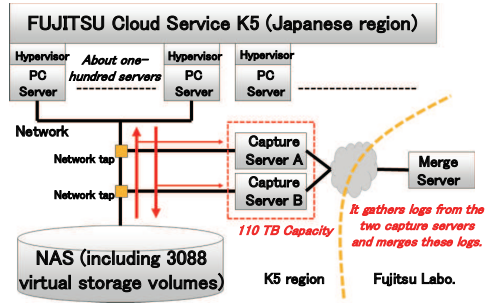}
\caption{Capture system for K5 cloud service}
\label{capture-system}
\end{center}
\vskip-\lastskip \vskip -5pt 
\end{figure}

The FUJITSU K5 cloud service operates in over 25 
data centers across four continents and includes more than 
5,300 cloud customers with over 10,000 expert cloud employees. 
It is also supported by more than 13,000 servers and 640 
business systems and is built using OpenStack technology, 
including VMware and Bare Metal stacks for specific workloads. 
It uses OpenStack technology to avoid vendor lock-in, and 
a user develops applications more easily by using pre-defined 
services with Cloud Foundry \cite{CL-FOUNDRY}. 
The K5 service consists of both Infrastructure as a Service (IaaS) 
and Platform as a Service (PaaS). 
We inserted some network taps \cite{NET-TAP} into the TCP/IP 
network between servers running on hypervisor and storage systems 
in a K5 data center in Japan \cite{WANC-2019} (See Figure 
\ref{capture-system}). 
The monitor's servers were operating some databases, mail server, 
and some Fujitsu internal services on IaaS.
We connected these network taps to a capture system. 
The capture system consisted of two servers and 110 TB storage. 
In order to analyze the captured data, we should gather the data 
from the capture system to the merge server installed in Fujitsu 
Labos. We also should merge these two data into one data.  
This merged data consisted of the timestamp, virtual storage volume 
ID, read/write, offset, and length information per request. 
%
Each virtual storage volume was assigned to a VM, and there were 
3088 workloads. 
%

%
%

%
We gathered these trace logs from Monday to Sunday but not 
continuously because one day's trace logs often consumed the storage 
capacity of the capture system and the merge functions spended 
more than two days. 
We divided these discontinuous seven days of trace logs by each 
virtual storage volume ID.
In this paper, we call this divided trace log a ``workload''. 
We retrieved the three sets of Monday-Sunday workloads (See 
Table \ref{LOG_DATE}) and analyzed the first set (dataset1) of 
these workloads. 
All the workloads already have been released from the IO trace web 
site of the Storage Networking Industry Association (SNIA) 
\cite{SNIA-IO-TRACE}. 

\begin{table}[tb] \caption{dates for the trace logs getted (2017/10/03--2018/03/29)} 
\label{LOG_DATE}
\begin{center}
\scriptsize
\begin{tabular}{|l|l|l|l|l|l|l|l|}
\hline
         & Sun   & Mon   & Tue   & Wed   & Thu   & Fri   & Sat \\ \hline
dataset1 & 11/26 & 10/30 & 10/03 & 11/08 & 10/12 & 10/20 & 11/18 \\ \hline
dataset2 & 01/07 & 12/04 & 01/16 & 01/24 & 01/16 & 12/15 & 12/23 \\ \hline
dataset3 & 02/25 & 03/05 & 03/13 & 03/21 & 03/13 & 02/16 & 02/10 \\ \hline
\end{tabular}
\end{center}
\end{table}

\section{Analysis of parallel workloads}

\label{ana_pw}

\subsection{How to analyze the parallel workloads}

\label{ana_pw_howto}

Our goal was to clarify how to build cost-effective hybrid 
storage system for cloud service platform. 
To achieve our goal, it was important for us to understand 
the access locality for the retrieved parallel workloads. 
Then, we analyzed these workloads from the viewpoint of both 
temporal and spatial access locality by using the dataset1 
of Table \ref{LOG_DATE}. 
First, we investigated the number of IO accesses among the 3088 
parallel workloads, and we selected the 38 parallel workloads 
because these workloads included half of all IO accesses for 3088 
workloads. 
We executed the rest analysis by using the 38 parallel workloads. 
Second, we analyzed temporal access locality to clarify the bias 
for IO access in the day. 
Third, we analyzed temporal and spatial access locality of a 
microscopic view. 
Microscopic view meant the analysis by page unit. 
This analysis was focused whether the traditional cahce replacement 
algorithms were effective or not for each workload.  
Last, we also analyzed temporal and spatial access locality of a 
macroscopic view. 
Macroscopic view meant the analysis by 1-GB unit.

\subsection{Selection of 38 parallel workloads}

\label{ana_pw_38}

\begin{figure}[tb]
\begin{center}
\includegraphics[width=8cm,height=4.5cm]{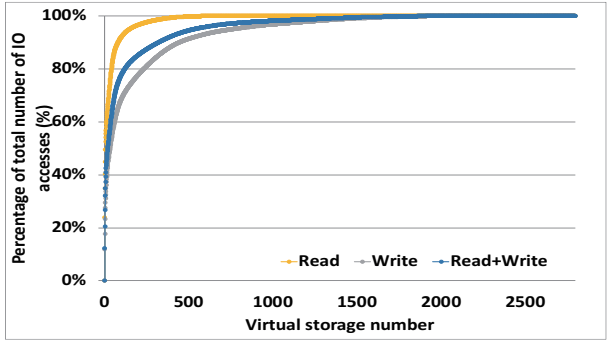}
\caption{Bias of number of IO accesses among 3088 workloads}
\label{bias-io}
\end{center}
\vskip-\lastskip \vskip -5pt 
\end{figure}

We first examined the bias of the number of read, write, and 
read+write IO accesses among the 3088 workloads and found that a small 
number of workloads included most IO accesses. 
Therefore, we selected 38 workloads whose percentage of the number of 
IO accesses was more than 50 with either read, write, or read+write and 
analyzed these 38 to quickly obtain both the temporal and spatial 
locality of all 3088 workloads 
(see Figure \ref{bias-io}). 
%

\begin{figure}[tb]
\begin{center}
\includegraphics[width=8cm,height=4.5cm]{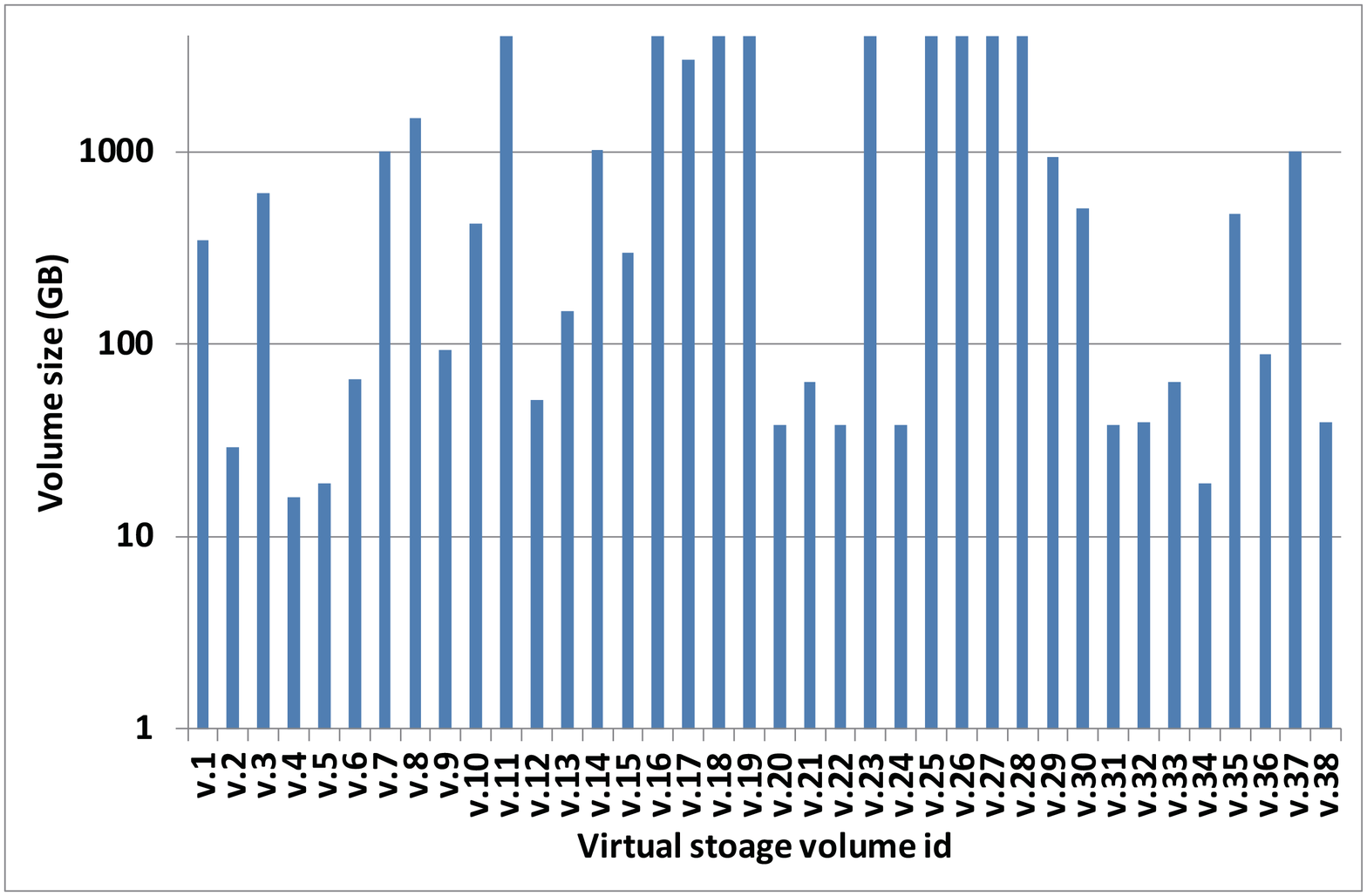}
\caption{Volume size (GB)}
\label{v_size}
\end{center}
\vskip-\lastskip \vskip -5pt 
\end{figure}

\begin{figure}[tb]
\begin{center}
\includegraphics[width=8cm,height=4.5cm]{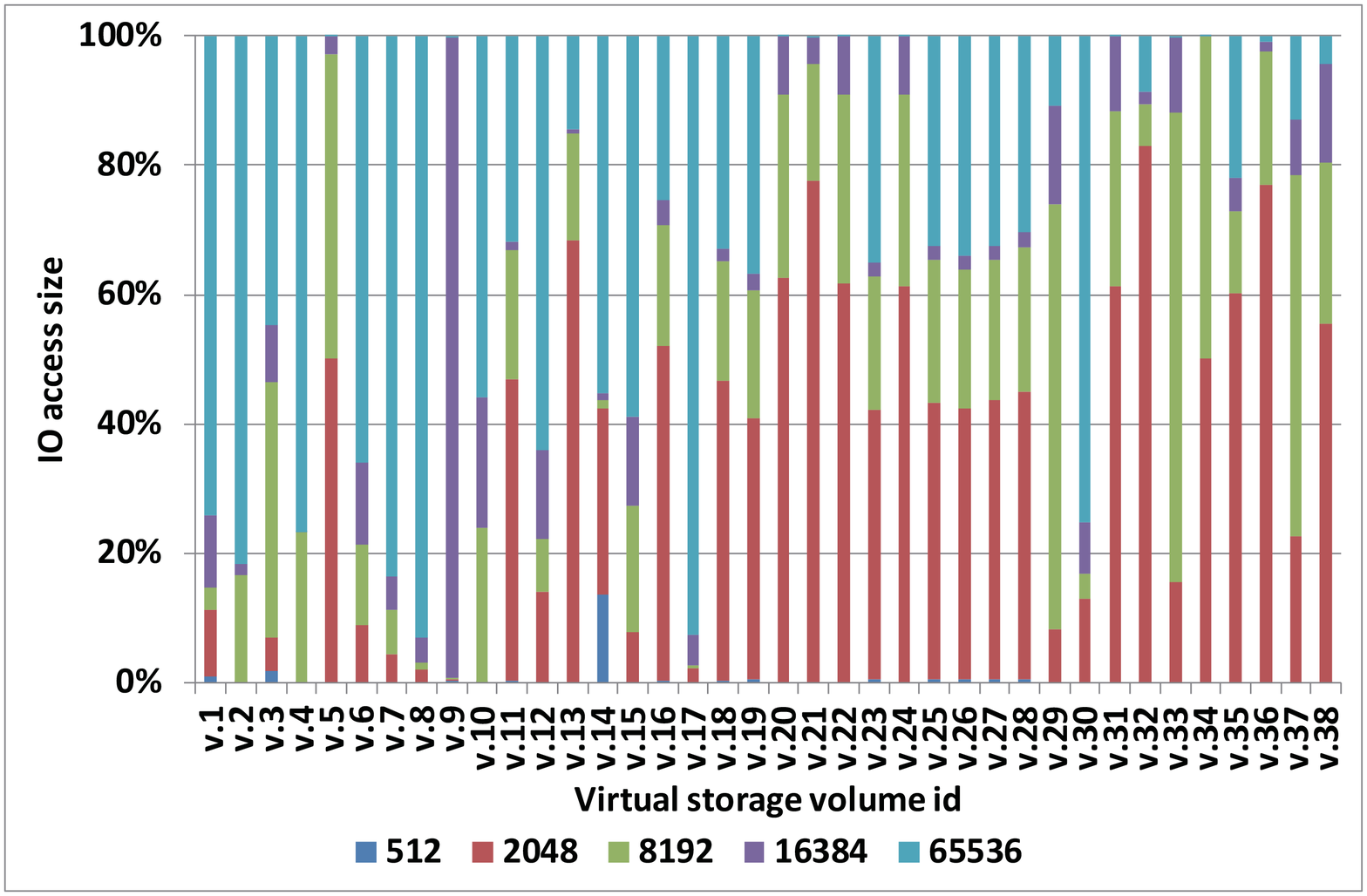}
\caption{IO access size}
\label{io_acc_size}
\end{center}
\vskip-\lastskip \vskip -5pt 
\end{figure}

Figure \ref{v_size} illustrates the volume sizes of the 38 workloads. 
We regarded the furthest footprint of IO access as the volume size 
of each workload because we could not obtain the volume size 
information. 
These volumes sizes were between 19 and 4000 GB. 
The x-axis of Figure \ref{v_size} shows the virtual storage 
volume IDs, which have almost the same mean as the workload. 
Figure \ref{io_acc_size} illustrates the IO access size of each 
virtual storage volume ID. 
The top ten workloads mainly had an IO size of more than 16 KB. 
However, the IO sizes of the remaining workloads were mainly less 
than 8 KB. 
%

%
%

%

\subsection{Analysis for temporal and spatial access locality}

\label{ana_pw_core}

\subsubsection{Temporal access locality}

\label{ana_pw_core_temp}

\begin{figure}[tb]
\begin{center}
\includegraphics[width=8cm,height=5cm]{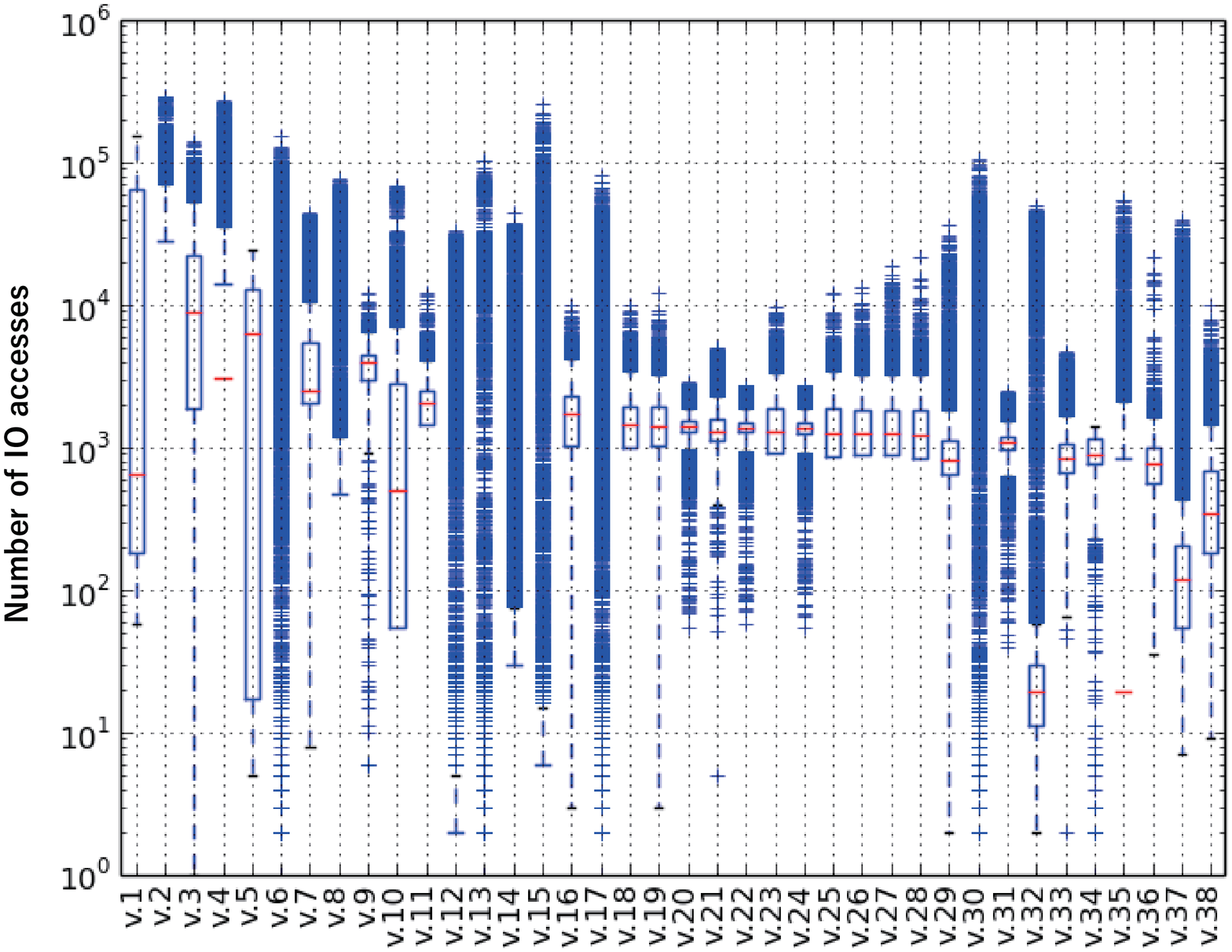}
\caption{Box-plot diagram for temporal access locality}
\label{ex2_3}
\end{center}
\end{figure} 
\begin{figure}[tb]
\begin{center}
\includegraphics[width=8cm,height=8cm]{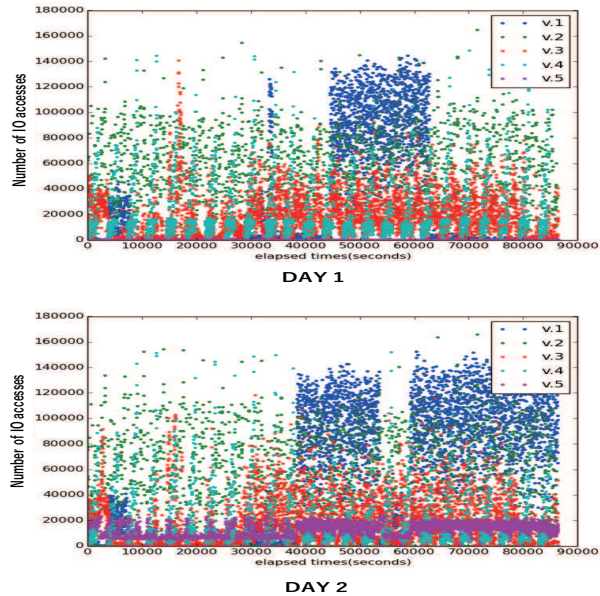}
\caption{Temporal access locality of top five workloads}
\label{ex2_1}
\end{center}
\vskip-\lastskip \vskip -5pt 
\end{figure} 
First, we analyzed the workloads for temporal access locality. 
We counted the number of IO accesses at 15-second intervals, saved 
these values, and created a box-plot diagram (Figure \ref{ex2_3}). 
The red line of Figure \ref{ex2_3} is the median, the top of Figure 
\ref{ex2_3}'s box is the upper quartile, and the bottom of Figure 
\ref{ex2_3}'s box is the lower quartile. 
We found that the number of IO accesses drastically varied for 
almost half the workloads, whereas it was almost stable for 
the remaining workloads. 
In order to show the actual access patterns, we choosed two day's 
data and analyzed the number of IO accesses for the top five 
workloads. 
Figure \ref{ex2_1} illustrated the results. 
We found that some workloads (especially v.1) drastically increased 
the number of IO accesses only at certain times. 
We also observed that these times changed depending on the day. 
On the other hand, the v.4 workload had a regular access pattern. 

\subsubsection{Temporal and spatial access locality of a microscopic view}

\label{ana_pw_core_temp_spa_micro}

\begin{figure}[tb]
\begin{center}
\includegraphics[width=8cm,height=5cm]{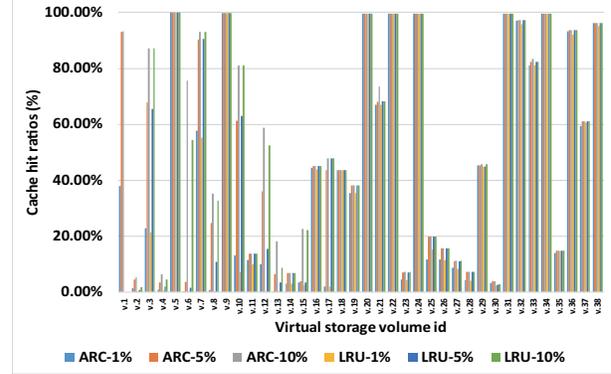}
\caption{Cache hit ratios (\%)}
\label{sim-ideal}
\end{center}
\vskip-\lastskip \vskip -5pt 
\end{figure} 
\begin{figure}[tb]
\begin{center}
\includegraphics[width=8cm,height=5cm]{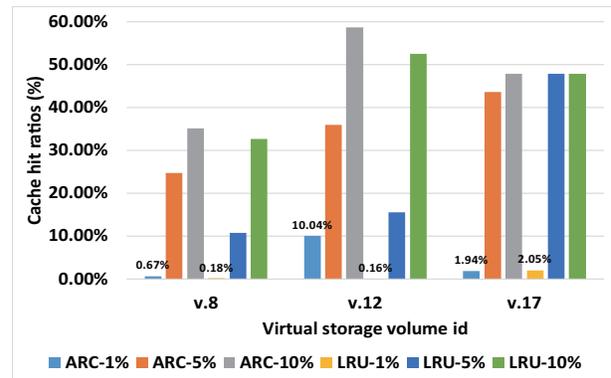}
\caption{Cache hit ratios (v.8, 12, and 17)(\%)}
\label{sim-ideal-2}
\end{center}
\vskip-\lastskip \vskip -5pt 
\end{figure} 
We investigated the temporal and spatial access locality 
of a microscopic view by using the sim-ideal cache 
simulator \cite{SIM-IDEAL}. 
The sim-ideal was developed at the University of Minnesota, and 
its default page-size was 4-KB.
We set 1, 5, and 10\% as each workload size for each workload's 
cache size. 
The cache replacement algorithms of the sim-ideal are Adaptive 
Replacement Cache (ARC) \cite{ARC-FAST2003} and Least Recently 
Used (LRU). 
We believe that ARC is the most effective cache replacement algorithm 
because it determines a replacement by using both recency and frequency. 
These algorithms have already been implemented in the sim-ideal. 
Figure \ref{sim-ideal} shows these results. 
These cache hit ratios varied with the workload, but almost half the 
workloads had low cache hit ratios (less than 20\%), and these ratios 
were almost the same even when we increased the cache size from 1 to 10\%. 
This means that these workloads included few 4-KB page level 
regularities and could not be handled effectively with LRU and ARC. 
Cache hit ratios of v.8, 12, and 17 increased according to the 
increase of the cache size (See Figure \ref{sim-ideal-2}). 
We should search their convergence points for cache hit ratios and 
set the corresponding cache size. For example, 5 \% was convergence 
point when v.17 was executing. 
Moreover, we should choose the more desirable cache replacement 
algorithm among the simulated algorithms if these cache hit ratios 
were differ. 
For example, when both v.8 and 12 were executing, the cache hit 
ratio of ARC was always better than that of LRU.

\subsubsection{Temporal and spatial access locality of a macroscopic view}

\label{ana_pw_core_temp_spa_macro}

\begin{figure}[tb]
\begin{center}
\includegraphics[width=8cm,height=5cm]{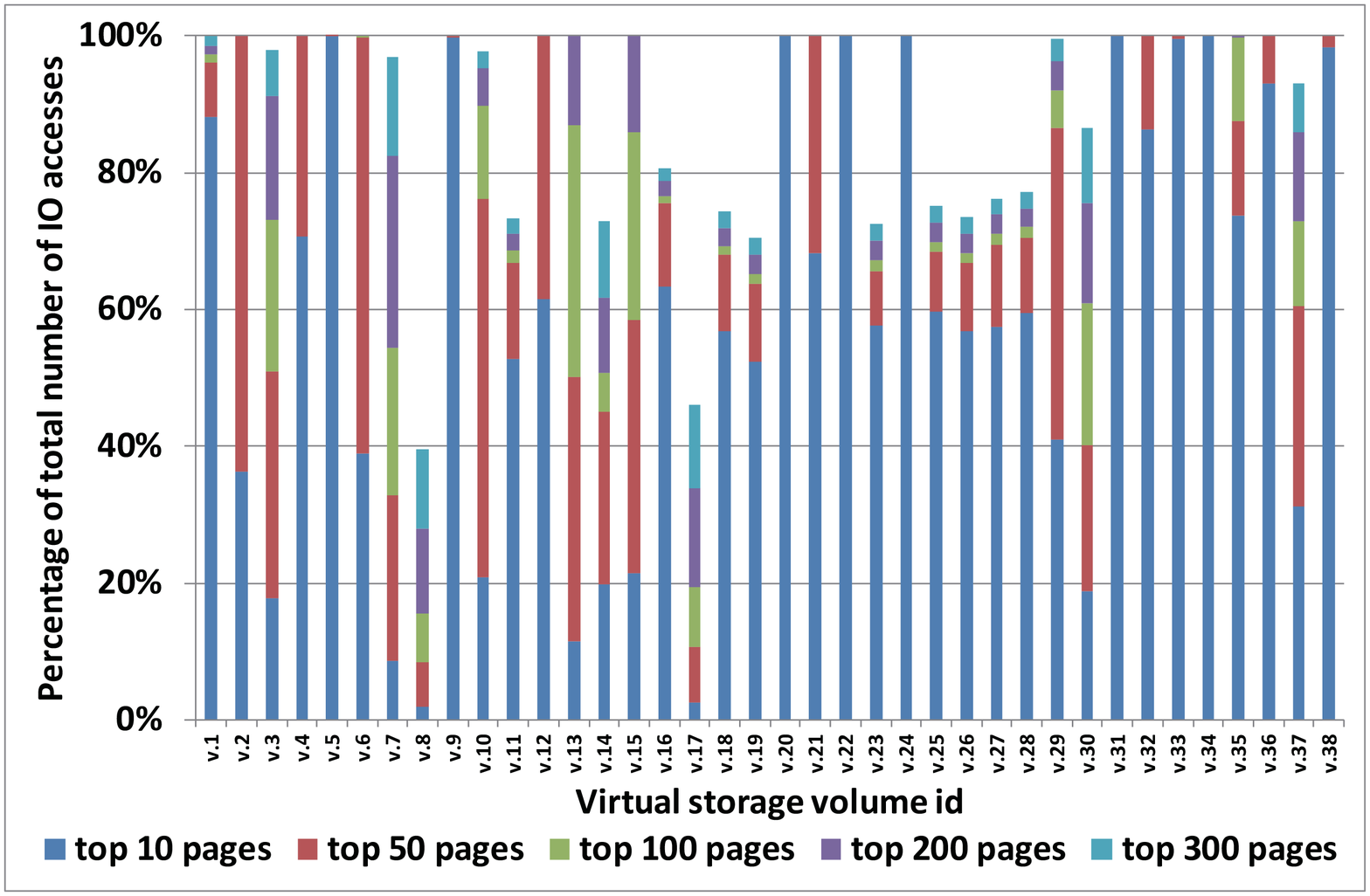}
\caption{Percentage of total number of IO accesses}
\label{ex5_1}
\end{center}
\vskip-\lastskip \vskip -5pt 
\end{figure} 
\begin{figure}[tb]
\begin{center}
\includegraphics[width=8cm,height=5cm]{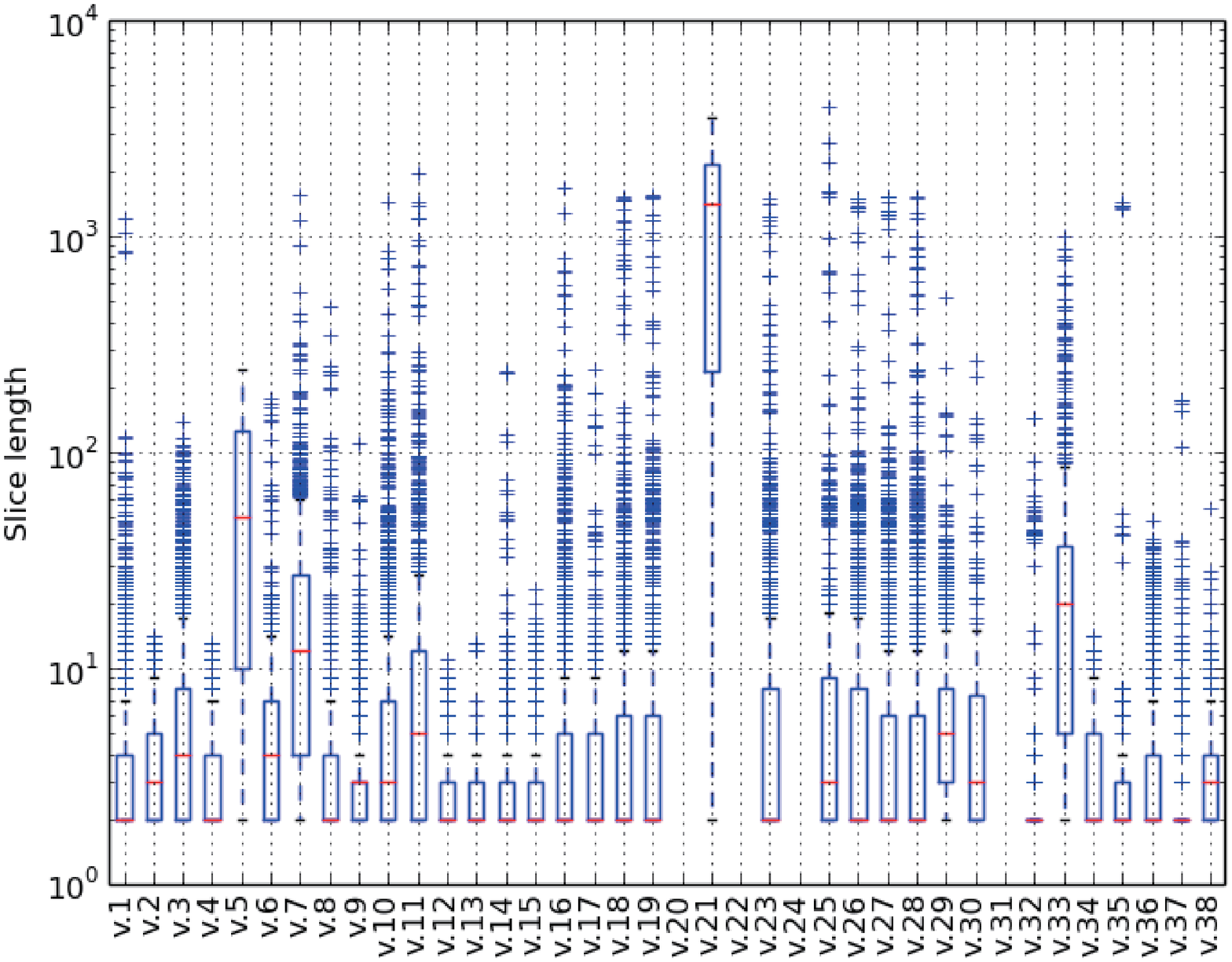}
\caption{Slice length (=Continuous times) for each workload}
\label{ex6_3}
\end{center}
\vskip-\lastskip \vskip -5pt 
\end{figure} 
\begin{figure}[tb]
\begin{center}
\includegraphics[width=8cm,height=5cm]{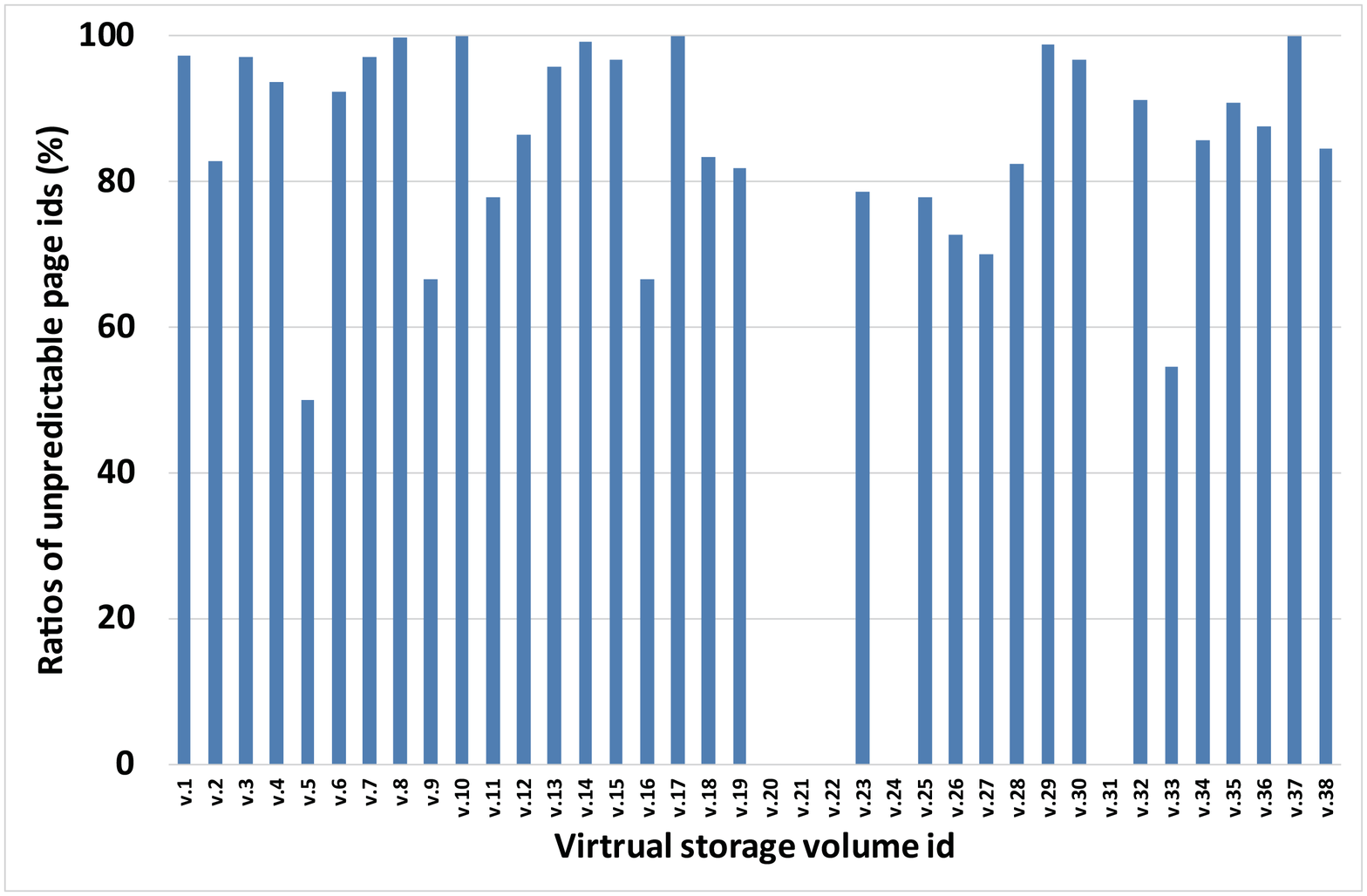}
\caption{Percentages of unpredictable page IDs}
\label{unpridictable_page_id}
\end{center}
\vskip-\lastskip \vskip -5pt 
\end{figure} 
We investigated temporal and spatial access locality 
of a macroscopic view. 
We then divided the workloads into 1-GB pages, counted the 
number of IO accesses to each page at 15-second intervals, 
sorted the pages in descending order of IO accesses at 15-second 
intervals, calculated the ratio to total IO accesses at these 
intervals, and averaged the ratios. 
Figure \ref{ex5_1} shows the results. 
We found that almost all workloads concentrated IO accesses in only 
a few pages because the top ten pages often included more than half 
the total number of IO accesses. 
We then investigated these concentrated IO accesses that occurred on 
consecutive pages. 
Figure \ref{ex6_3} illustrates the duration of the above concentrated 
IO-access pages. 
We now explain how we retrieved the concentrated IO-access pages. 
First, we sorted pages in descending order of IO accesses and 
integrated the number of IO accesses in order of the sorted pages. 
If the integrated number of IO accesses was more than half the total 
number of IO accesses, we stopped the above integration and selected the 
pages used for this integration. 
These concentrated IO accesses continued on consecutive pages for 
more than 15 seconds because one slice on the y-axis is 15 seconds. 
Moreover, some of these concentrated IO accesses continued for 
more than 150 seconds.
Some values from the upper quartile to the maximum continued for more 
than 1500 seconds.
Next, we investigated the repeatability of the concentrated IO access 
pages. 
We now explain how we determined repeatability. 
First, we counted the number of concentration judgments per page 
and calculated the ratio of all judgments per page. 
If this ratio was less than 5\%, a page was judged as 
unpredictable. 
Figure \ref{unpridictable_page_id} shows the results. 
Almost all workloads had more than 50\% unpredictable 
pages. 
Therefore, it was difficult for us to predict the appearance of 
the concentrated IO access pages. 
%

The results of these concentrated IO access pages were almost the 
same as those of IO concentration, which was shown in already 
published documents \cite{EAIS-2016,IEICE-D-12}. 

\section{How to handle parallel workloads for hybrid storage systems}

\label{handle_pw}

Section \ref{ana_pw_38} shows that IO accesses for only 38 workloads 
included more than 50\% of IO accesses for 3088 workloads. 
Therefore, a perferable hybrid storage system will carefully select the 
workloads. 
Specifically, it should assign the regions for first tier only when 
a workload includes a large number of IO accesses. 
Section \ref{ana_pw_core_temp} shows that the number of IO accesses 
drastically varied for almost half the workloads. 
Figure \ref{ex2_1} also shows that some workloads drastically 
increased the number of IO accesses only at certain times. 
Therefore, a preferable hybrid storage system will monitor a change 
in IO accesses for a short interval. 
It should dynamically assign the regions for first tier if the number 
of IO accesses for a workload is drastically increased. 
Section \ref{ana_pw_core_temp_spa_micro} shows that almost half the 
workloads had a tiny effect on applying caching. 
A preferable hybrid storage system then should use two approaches for 
handling these workloads. 
The first approach is to assign a whole workload area to regions 
for first tier if one would like to improve the performance of these 
workloads. 
The second is not to assign first tier's regions for the workload if 
one would not like to improve the performance. 
Some cache hit ratios (e.g., v.8, 12, 17) increased according to the 
increase in the cache size. 
Their convergence points for cache hit ratios should be searched for, 
and the corresponding cache size should be set. 
For example, 5\% was the convergence point when v.17 was executing. 
A more preferable cache replacement algorithm should be chosen among 
the candidate algorithms if their cache hit ratios differ. 
For example, when both v.8 and 12 were executing, the cache hit 
ratio of ARC was always better than that of LRU.
Section \ref{ana_pw_core_temp_spa_macro} shows that concentrated 
IO accesses often occurred on a 1-GB page unit, and these features 
were almost the same as those of IO concentration. 
IO concentration targets narrow regions of storage volume and can 
continue for up to an hour. 
These narrow regions occupy a few percent of the logical unit 
number capacity, are the target of most IO accesses, and appear 
at unpredictable logical block addresses. 
ATSMF \cite{IEICE-D-12} is a hybrid storage system for effectively 
handling IO concentration. 
We will check whether ATSMF is adequate for these parallel 
workloads in the near future. 

\section{Related work}

Sundaresan et al. \cite{IC2E_2016} proposed Multi-Cache, a 
multi-layer cache management system that uses a combination of 
cache devices of varied speed and cost such as SSDs and 
non-volatile memories (NVMs), to dynamically allocate cache 
capacities among different VMs. 
Multi-Cache partitions each device dynamically at runtime 
in accordance with the workload of each VM and its priority. 
It uses a heuristic optimization technique that ensures the 
maximum utilization of caches, resulting in a high hit ratio. 
We think that Multi-cache can improve its performance by 
using the knowledge of this research. 
We also showed previous workload studies for IO concentration 
because the results for Section \ref{ana_pw_core_temp_spa_macro} 
indicated some characteristic for IO concentration. 
IO concentration often included various applications on shared 
file systems, mail servers, and web servers 
\cite{OS-2012-12,WRITE-OFF-LOADING,MSR-Cambridge,EAIS-2016}. 
IO concentration appeared in narrow regions of a storage volume 
and continued for durations of up to about an hour. 
These narrow regions occupied only a small percentage of the 
storage volume and either remained for a long time or shifted 
to a neighboring area of the storage volume after several minutes 
on average. Furthermore, these narrow regions included most 
IO accesses and appeared in unpredictable logical block addresses 
(LBAs).

\section{Conclusion}

\label{conc}

We retrieved and analyzed parallel storage workloads of the FUJITSU K5 
cloud service to clarify how to build cost-effective hybrid storage 
systems for cloud service platform. 
A hybrid storage system consists of fast but low-capacity 
tier (first tier) and slow but high-capacity tier (second tier). 
And, it typically consists of either SSDs and HDDs or NVMs and 
SSDs.  
We then investigated these workloads from the viewpoint of both 
temporal and spatial access locality and had the following ideas 
for building a cost-effective hybrid storage system. 
A preferable hybrid storage system should dynamically optimized across 
workloads by using the following ideas because multiple workloads with 
different characteristics were running in parallel. 

\begin{itemize}
\item Assign regions for first tier only if a workload includes a 
large number of IO accesses for a whole day because some workloads 
may include a tiny number of IO accesses. 
\item Dynamically choose the regions that include a large number 
of IO accesses and move them from second tier to first tier for a 
short interval because some workloads drastically increase the number 
of IO accesses only at certain times. 
\item Check the cache hit ratio of each workload. If a cache hit ratio 
is regularly low, the use of cache for the workload should be 
cancelled, and the whole workload region should be assigned to the 
region for first tier. 
If a cache hit ratio increases according to the increase of the 
cache size, the convergence point for the cache hit ratio should 
be searched for, and the corresponding cache size should be set. 
If the cache hit ratios differ among the candidate cache replacement 
algorithms, a more preferable algorithm should be chosen. 
\item Use the features for concentrated IO accesses on a narrow 
region. These features are almost the same as those of IO 
concentration. ATSMF is one candidate system because it can 
effectively handle IO concentration. 
\end{itemize}

\section{Future work}

To be satisfied the ideas described in Section \ref{conc}, 
the preferable hybrid storage system should dynamically analyze the 
temporal and spatial access locality of workloads with many IO 
accesses, decide the preferable cache replacement algorithm including 
ATSMF and capacity of caching for each workload, and replace the 
decided cache replacement algorithm and capacity of first tier from 
the previous algorithm and capacity for each workload. 
We will implement and evaluate this hybrid storage system in the 
near future. 

{\footnotesize \bibliographystyle{acm}
\bibliography{k5_analysis}}

\begin{thebibliography}{10}

\bibitem{CL-FOUNDRY}
{Cloud Foundry}.
\newblock {https://en.wikipedia.org/wiki/Cloud\_Foundry}.

\bibitem{FUJITSU_K5}
{Fujitsu Cloud Service K5}.
\newblock {http://www.fujitsu.com/global/services/hybrid-cloud/k5/}.

\bibitem{NET-TAP}
{Network tap}.
\newblock {https://en.wikipedia.org/wiki/Network\_tap}.

\bibitem{OpenStack}
{OpenStack}.
\newblock {https://en.wikipedia.org/wiki/OpenStack}.

\bibitem{SNIA-IO-TRACE}
{Parallel Traces for SNIA IOTTA Repository}.
\newblock {http://iotta.snia.org/tracetypes/4}.

\bibitem{SIM-IDEAL}
{sim-ideal}.
\newblock {https://github.com/arh/sim-ideal}.

\bibitem{MSR-Cambridge}
{SNIA trace data MSR Cambridge}.
\newblock {http://iotta.snia.org/traces/388}.

\bibitem{ARC-FAST2003}
{\sc Megiddo, N., and Modha, D.~S.}
\newblock {ARC: A SELF-TUNING,LOW OVERHEAD REPLACEMENT CACHE}.
\newblock In {\em Proc. 2th USENIX conference on File and Storage Technorogies
  (FAST2003)\/} (March 2003).

\bibitem{WRITE-OFF-LOADING}
{\sc Narayanan, D., Donnelly, A., and Rowstron, A.}
\newblock {Write Off-Loading: Practical Power Management for Enterprise
  Storage}.
\newblock In {\em in Proc. of 6th USENIX Conf. on File and Storage Tech\/} (Feb
  2008).

\bibitem{OS-2012-12}
{\sc Oe, K., Honda, T., and Kawaba, M.}
\newblock {Samba workload analysis and consideration for hybrid storage
  system}.
\newblock In {\em IPSJ SIGOS, Tokyo, Japan (in Japanese)\/} (Dec 2012).

\bibitem{EAIS-2016}
{\sc Oe, K., Nanri, T., and Okamura, K.}
\newblock {Analysis of storage workloads of input-output access locality and
  designing of hybrid storage system}.
\newblock In {\em Proc. of the 5th IIAI International Congress on Advanced
  Applied Informatics\/} (July 2016).

\bibitem{IEICE-D-12}
{\sc Oe, K., Sato, M., and Nanri, T.}
\newblock {ATSMF: Automated Tiered Storage with Fast Memory and Slow Flash
  Storage to Improve Response Time with Concentrated Input-Output (IO)
  Workloads}.
\newblock {\em IEICE Transactions on Information and Systems (VOL.E101-D
  No.12)\/} (December 2018).

\bibitem{WANC-2019}
{\sc Ogihara, K.}
\newblock {Generating block IO trace data from a cloud site using packet
  capture and analyzing the IO trace data}.
\newblock In {\em 10th International Workshop on Advances in Networking and
  Computing (WANC'19), Nagasaki, Japan\/} (Nov 2019).

\bibitem{IC2E_2016}
{\sc Rajasekaran, S., Duan, S., Zhang, W., and Wood, T.}
\newblock {Multi-Cache: Dynamic, Efficient Partitioning for Multi-Tier Caches
  in Consolidated VM Environments}.
\newblock In {\em Proc. of Cloud Engineering (IC2E), 2016 IEEE International
  Conference, Berlin, Germany\/} (April 2016).

\end{thebibliography}



\end{document}